# Giant Magnetoelectric Effect in a Multiferroic Material with a High Ferroelectric Transition Temperature

N. Hur,[1] I. K. Jeong,[2] M. F. Hundley,[3] S. B. Kim,[4] S.-W. Cheong[4]

[1]Department of Physics, Inha University, Incheon 402-751, Korea

[2]Research Center for Dielectrics and Advanced Matter Physics, Pusan National University, Busan 609-735, Korea

[3]Los Alamos National Laboratory, Los Alamos, NM 87545, USA.

[4]Rutgers Center for Emergent Materials, Department of Physics and Astronomy, Rutgers University, Piscataway, NJ 08854, USA.

We present a unique example of giant magnetoelectric effect in a conventional multiferroic $HoMnO_3$, where polarization is very large (~56 $mC/m^2$) and the ferroelectric transition temperature is higher than the magnetic ordering temperature by an order. We attribute the uniqueness of the giant magnetoelectric effect to the ferroelectricity induced entirely by the off-center displacement of rare earth ions with large magnetic moments. This finding suggests a new avenue to design multiferroics with large polarization and higher ferroelectric transition temperature as well as large magnetoelectric effects.

Magnetoelectric multiferroics had been an interesting research field for several decades [1-4] and they were recently revived actively due to discoveries of spectacular magnetoelectric (ME) or magnetodielectric response in several materials [5-13] including orthorhombic manganites ((Gd, Tb, Dy)$MnO_3$, (Tb, Dy)$Mn_2O_5$), Kagome-staircase $Ni_3V_2O_8$, and hexaferrite $Ba_{0.5}Sr_{1.5}Zn_2Fe_{12}O_{22}$. Common features of these new materials are low ferroelectric transition temperature ($T_C$), small spontaneous polarization ($P$) (of the order of 100 µC m$^{-2}$), and strong ME effects. Recent theoretical studies proposed that the ferroelectricity in these materials is induced by magnetic ordering of peculiar structures, such as noncolinear or long-wavelength magnetic structure [15, 16]. Despite strong ME effects, magnetically driven ferroelectrics are not so convenient for practical applications in some sense because they have to be poled frequently by applying an electric field to maintain the single ferroelectric domain state. This is because they have too small polarization which can be easily broken into multi-domain polarization and also because the ME effect arises mainly from the phase transition between the paraelectric state and the ferroelectric state [13, 14]. Thus, multiferroics with a high $T_C$ and a robust large polarization are more desirable because they easily maintains the single ferroelectric domain state once they are poled, exhibiting ME effect just by applying $H$, which makes them much convenient for the real application.

Traditional way to achieve multiferroics with large $P$ and high $T_C$ is to search for conventional ferroelectrics containing magnetic ions. For example, lone-pair driven multiferroic $BiMnO_3$ ($BiFeO_3$) [17], geometrically driven multiferroic $YMnO_3$ [18], and $BaMnF_4$ ($BaCoFe_4$) [4] have high $T_C$ ranging from 700 K to 1000 K and $P$ larger than 10 mC m$^{-2}$. However, in these 'high $T_C$ multiferroics', no examples are known that show large ME effects comparable to those of magnetically driven multiferroics with low $T_C$ in moderate magnetic fields ($H$) of a few tesla or less. These high $T_C$ multiferroics usually exhibit just slight anomalies in dielectric constants ($\varepsilon$) at $T_M$ and show negligible $H$ dependence of $\varepsilon$ or $P$ [4, 17, 18]. In this letter, we demonstrate a drastic change in $P$ as well as $\varepsilon$ at the magnetic transition temperatures and giant ME effects in a high $T_C$ multiferroic $HoMnO_3$. Our crystallographic study suggests that the rare earth Ho ions are wholly responsible for the ferroelectricity as in the case of $YMnO_3$ [19]. We discussed the origin of the giant ME effect with the competing mechanism between the in-plane and inter-plane magnetic exchange interactions among Mn and Ho moments. Our study provides an interesting model for the behavior of polarization when the antiferromagnetic exchange interaction among the moments of ferroelectric active ions is triggered along the polarization direction. In addition, our result offers a clear-cut explanation for the $\varepsilon$ anomalies at the magnetic transitions, of which the origin has been under debate [20, 21].

Extensive studies have been recently performed on hexagonal manganites, especially on $HoMnO_3$, due to its rich phase diagram and the strong spin-lattice coupling [20, 22, 23]. However, any direct observation on the evolution of the spontaneous $P$ (which is in fact one of the most important order parameters in multiferroics) has not been made. $HoMnO_3$ exhibits ferroelectric ordering with $T_C \approx 900$ K and antiferromagnetic (AFM) ordering at $T_N \approx 70$ K [11, 18]. Owing to the detailed studies in optical second harmonic generation (SHG) [11, 21, 24, 25] and the neutron scattering [26-28], the magnetic structure had been well established. Thus, $HoMnO_3$ is a prototypical system in which one can study the effect of magnetic structures on the dielectric properties and the electric polarization.

In Fig. 1(a), we show $P$ and $\varepsilon$ along the $c$ axis as a function of temperature ($T$) for HoMnO$_3$ single crystal that was grown using an optical floating zone furnace. The long range antiferromagnetic Mn$^{3+}$ ordering in the $ab$ plane at ~72 K is evidenced by the slight anomalies both in $\varepsilon$ and $P$. The slight kink in $P$ can be associated with the overall expansion of the lattice along the $c$ axis through the indirect spin-lattice coupling mechanism [23]. Upon further cooling, HoMnO$_3$ exhibits successive spin reorientation transitions at $T_{SR}$ and $T_{Ho}$ (Fig. 1(a)). It has been known by optical SHG and the neutron scattering that Mn spins are reoriented in plane by 90° at $T_{SR} \approx 38$ K and Ho spins orders antiferromagnetically along the $c$ axis at $T_{Ho} \approx 5$ K [24, 28]. The most striking feature in Fig. 1(a) is the drastic change in $P$. $P$ decreases abruptly at $T_{SR}$ and increases back again at $T_{Ho}$ upon cooling with a lower value of $P$ in the P6´$_3$cm´ phase.

Coupling between magnetic and ferroelectric orders is more clearly manifested in Figs 1(b) and 1(c). As $H$ increases, $T_{SR}$ significantly shifts to the lower $T$ and the lower boundary of the P6´$_3$cm´ phase moves to the higher $T$; consequently the P6´$_3$cm´ phase shrinks and completely disappears at ~5 T. We discovered that these $\varepsilon$ anomalies are intimately associated with the $P$ change. $P$ demonstrates abrupt change at the $T$ where $\varepsilon$ shows a sharp peak, and as the peak in $\varepsilon$ broadens out in $H$, the step-like change in $P$ also becomes broad and disappear above 5 T. There have been a few explanations for the increase of $\varepsilon$ in the intermediate phase (INT phase) between P6´$_3$cm´ and P6´$_3$c´m: a $z$-$z$ ME coupling of magnetization ($M$) and $P$ enabled by the Mn spin canting along the $z$ axis [20], the effect of unpinning between ferroelectric and AFM domain walls [20], and the pronounced local ME effect due to massive formation of AFM domain walls observed by magneto-optical SHG [21]. Our polarization data measured using a poled HoMnO$_3$ crystal with single ferroelectric domain clearly indicate that the $\varepsilon$ enhancement originates from the quasi-static change of $P$ across the INT phase rather than transient effects confined within ferroelectric or AFM domain walls. Another noticeable thing in Fig. 1(c) is that the magnitude of total change in $P$ (~ 80 µC m$^{-2}$) is almost comparable to the giant ME effect (order of 100 µC m$^{-2}$) observed recently in magnetically induced multiferroics [13, 14].

The evolution of ferroelectric properties across magnetic phases and the relation between $\varepsilon$ and $P$ are more clearly displayed in Fig. 2(a)-(d). First, at 18 K, as expected from Fig. 1(b) and 1(c), $\varepsilon(H)$ increases sharply at ~3.7 T and decreased again at ~4.5 T with a plateau structure in the INT phase, accordingly $P(H)$ increases linearly in the INT phase. It should be noted that no appreciable ME effects are observed either in the low field P6´$_3$cm´ or in the high field P6´$_3$c´m phase. However, a pronounced linear ME effect (5) (linear ME coefficient, $\alpha_{33}$, of ~59 ps m$^{-1}$) is observed in the INT phase, which has been assumed to be of P6´$_3$ symmetry where the Mn spins rotate to an angle between 0° and 90° with respect to the $a$ axis [20]. In fact even in this P6´$_3$ symmetry, a direct ME coupling between $P$ and the in-plane Mn moment is not allowed [1]. Thus, this suggests that the change in $P$ may be caused via a $H$-induced phase transition across a broad phase boundary (or P6´$_3$ phase) rather than via a linear ME effect in a specific symmetry that allows it.

Below T$_{Ho}$, $\varepsilon$ and $P$ show a more complicated field dependence. In Fig. 2(e), we summarized the $H$ dependence of $\varepsilon$ and constructed a $T$ versus $H$ phase diagram. About 5 phase boundaries were identified by peak, deep, or step-like $\varepsilon$ anomalies, which is pretty consistent with previous studies [22, 26]. Complicated $H$-dependence of $\varepsilon$ below T$_{Ho}$ is also accompanied by a drastic change in $P$ as shown in Fig. 2(b)-(d). Below 3 K, $P$ changes non-monotonically as a function of $H$ with the total change of $P$ as large as ~58 µC m$^{-2}$ at 2.5 K. The magnetic symmetry of the low-$T$ high-field (LTHF) phase is also still in question. There has been a proposition that the LTHF phase may be in $P6_3c'm'$ symmetry that allows linear ME effect [24]. However, no measurable linear ME effect was detected in our $P$ versus $H$ data at 3 K and 2.5 K in high field region (> 3 T) as shown in Fig. 2(c) and 2(d).

In order to resolve the origin of the gigantic ME effect, we performed the neutron powder diffraction study on HoMnO$_3$ in low $T$ region. Although any noticeable crystallographic anomalies were not detected across magnetic transitions (this is generally expected because the value of $P$ change in multiferroics usually corresponds to an average ionic displacement of the order of 10$^{-4}$ Å), our study on the ionic structure enabled us to discuss the mechanism of the giant ME effect in terms of the origin of ferroelectricity and the magnetic exchange interaction among Mn and Ho moments. Table 1 shows atomic parameters for HoMnO$_3$ refined from neutron powder diffraction experiment at 40 K. Fig. 3 also displays the crystal structure based on our atomic parameters and the magnetic structure of HoMnO$_3$ for two magnetic symmetries. Recent studies have provided experimental and theoretical evidences that the electric polarization in hexagonal YMnO$_3$ is originated from a buckling of MnO$_5$ polyhedra and vertical shift of the Y ions with Mn ions remaining very close to the center of the oxygen bipyramids [19, 29]. Our crystallographic analysis also showed that the vertical shift of rare earth Ho ions is wholly responsible for the ferroelectricity in HoMnO$_3$ unlike the previous experimental study that indicated Mn ions considerably contribute to $P$ by having different bond lengths between Mn and two apical oxygens in the bipyramid [27]. Our result points out that the giant change of $P$ as a function of $T$ and $H$ should be related to the magnetic ordering of Ho ions that are directly responsible for the occurrence of ferroelectricity.

Upon cooling, at $T_{SR}$, the d-f exchange interaction between Mn and Ho moments induces ordering of Ho moments in $P6'_3cm'$ phase as shown in Fig. 3 [11, 28]. Ho1 and Ho2 moments order along the $z$ with a ferrimagnetic arrangement in the $x$-$y$ plane leaving uncompensated moments. Thus, in this temperature range, the magnetic exchange interactions are predominantly along the $z$ axis through the Mn-Ho and the uncompensated inter-plane Ho AFM exchange interaction. This is strongly supported by the recent thermal expansion data which showed lattice shrinkage at the magnetic transition $T$ upon cooling along the direction of magnetic exchange interaction to gain exchange energy [23]. The lower value of $P$ in the $P6'_3cm'$ phase suggests that the strong AFM interaction along the $z$ accompanies a slight shift of ferroelectric active ions toward the centrosymmetric position. Note that the overall $c$-axis shrinkage at $T_{SR}$ ($\Delta c/c \approx -2.4\times10^{-6}$) is too small to be responsible for the polarization change ($\Delta P/P \approx -10^{-3}$) [23]. Upon further cooling, at $T_{Ho}$, the in-plane AFM ordering of Ho moments in $P6_3cm$ phase is triggered through the shorter Ho2-Ho2 exchange path (Ho1-Ho2$\approx$3.56 Å, Ho2-Ho2$\approx$3.53 Å) [11, 28]. The Ho1 moments surrounded by equal number of up and down moments of Ho2 ions, remain disordered. This onset of strong in-plane AFM Ho ordering, leaving no uncompensated in-plane moments, weakens the inter-plane exchange interaction along the $z$ axis, which again allows $P$ to restore its higher value in the $P6_3cm$ phase. This proposed mechanism of the competing in-plane and inter-plane exchange interactions, given by the peculiar crystallographic constraint, may explain the dome-like phase diagram of P6´$_3$cm´ phase which becomes most stable at ~20 K and shrinks back as the competing in-plane Ho ordering starts to be effective and thus helps the suppression of the P6´$_3$cm´ phase at lower $H$ in low $T$ region as shown in Fig. 2(e).

In summary, we observed giant ME effect in a high $T_C$ multiferroic HoMnO$_3$, which is intimately related to the off-center displacement of ferroelectric active Ho ions with very large moment (the largest total angular moment J=8 among rare earth ions). This indicates that the highly anisotropic 4f-electrons tightly coupled with the large rare earth moment can play a crucial role in provoking ME effect even in conventional high $T_C$ multiferroics.

Authors thank T. Kimura for help with measurements. Work at Los Alamos National Laboratory was performed under the auspices of the United States Department of Energy.

**Figure and Table Captions**

FIG. 1 (color online). (a) $T$ dependence of $P$ (solid circle) and $\varepsilon$ (open circle) for HoMnO$_3$ in zero $H$. Vertical dashed lines indicate magnetic phase boundaries. (b) $T$ dependence of $\varepsilon$ along the $c$ axis in various $H$ applied along the $c$ axis. (c) $T$ dependence of $P$ along the $c$ axis in $H$. $\varepsilon$ was measured with a 1 kHz ac electric field applied along the $c$ axis. $P$ was calculated by integrating the pyroelectric current. Before pyroelectric current measurement, the sample was cooled down from 900 K to 3 K in a varying electric field of 1-10 kV cm$^{-1}$ along the $c$ axis.

FIG. 2 (color online). $H$ dependence of the field-induced $P$ and the $\varepsilon$ change at (a) 18 K; (b) 4.5 K; (c) 3.0 K; (d) 2.5 K. The field-induced $P$ was obtained by measuring the magnetoelectric current as a function of $H$, which was varied linearly with time at the uniform rate of 100 Oe s$^{-1}$. (e) $T$ and $H$ phase diagram for HoMnO$_3$ overlaid with $\varepsilon$ change versus $H$ for each $T$. Dashed lines indicate phase boundaries and hatched regions are the intermediate phase (> 5 K) and the field hysteresis region (< 5 K). Note that the $T$ scale is expanded below 5 K.

TABLE 1. Atomic parameters for HoMnO$_3$ refined from neutron powder diffraction experiment at 40 K within the hexagonal space group P6$_3$cm. Refined lattice parameters are a=6.1173(1) Å and c=11.4107(2) Å.

FIG. 3 (color online). The crystallographic and magnetic structure of HoMnO$_3$ for two magnetic symmetries, P6´$_3$cm´ and P6$_3$cm. Crystallographic structure is based on the atomic parameters from our neutron powder diffraction experiment. Magnetic structures are adopted from [11, 28]. Arrows indicate magnetic moments of Ho ion. Moments of Mn ions are not displayed.

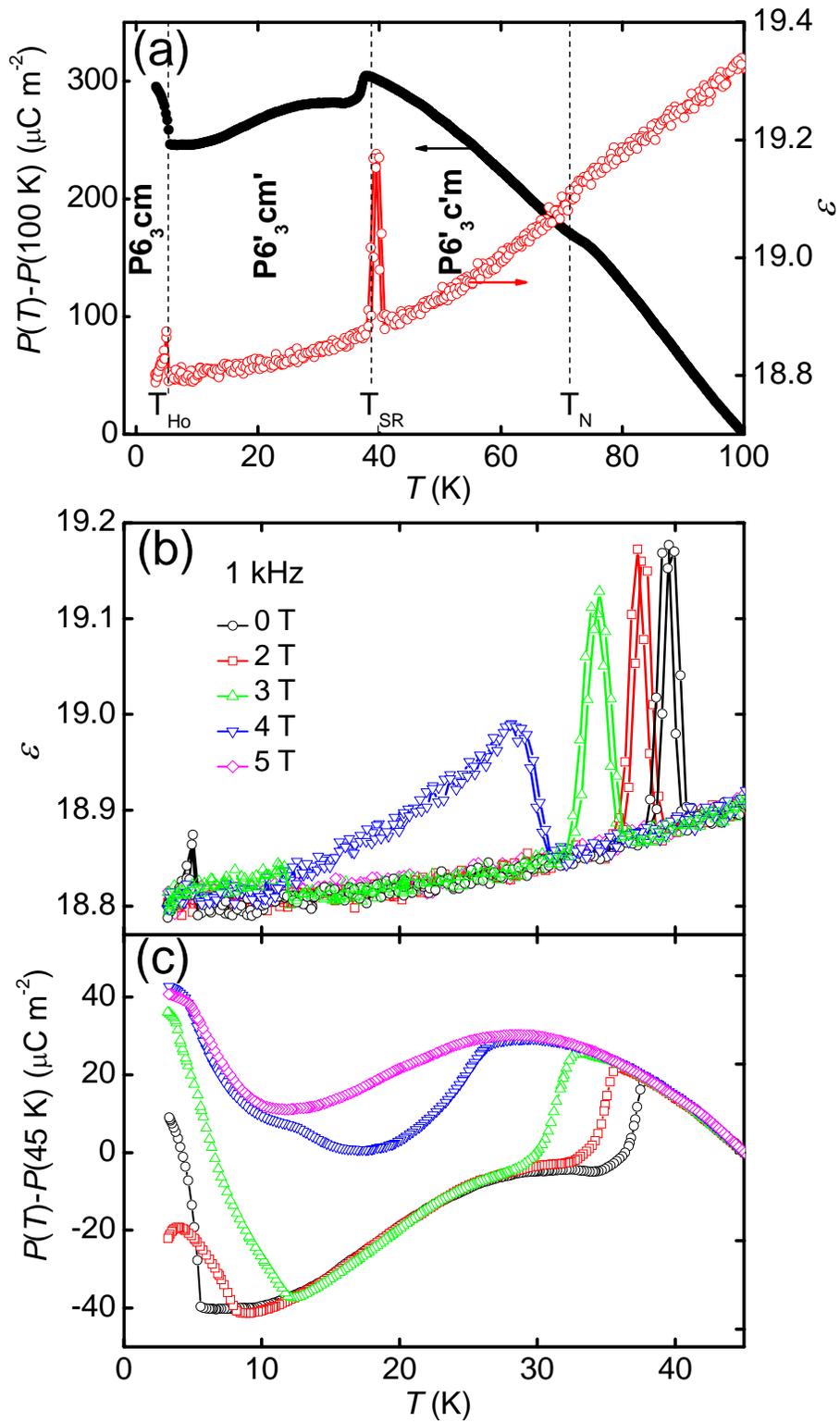

Fig. 1 N. Hur *et al.*

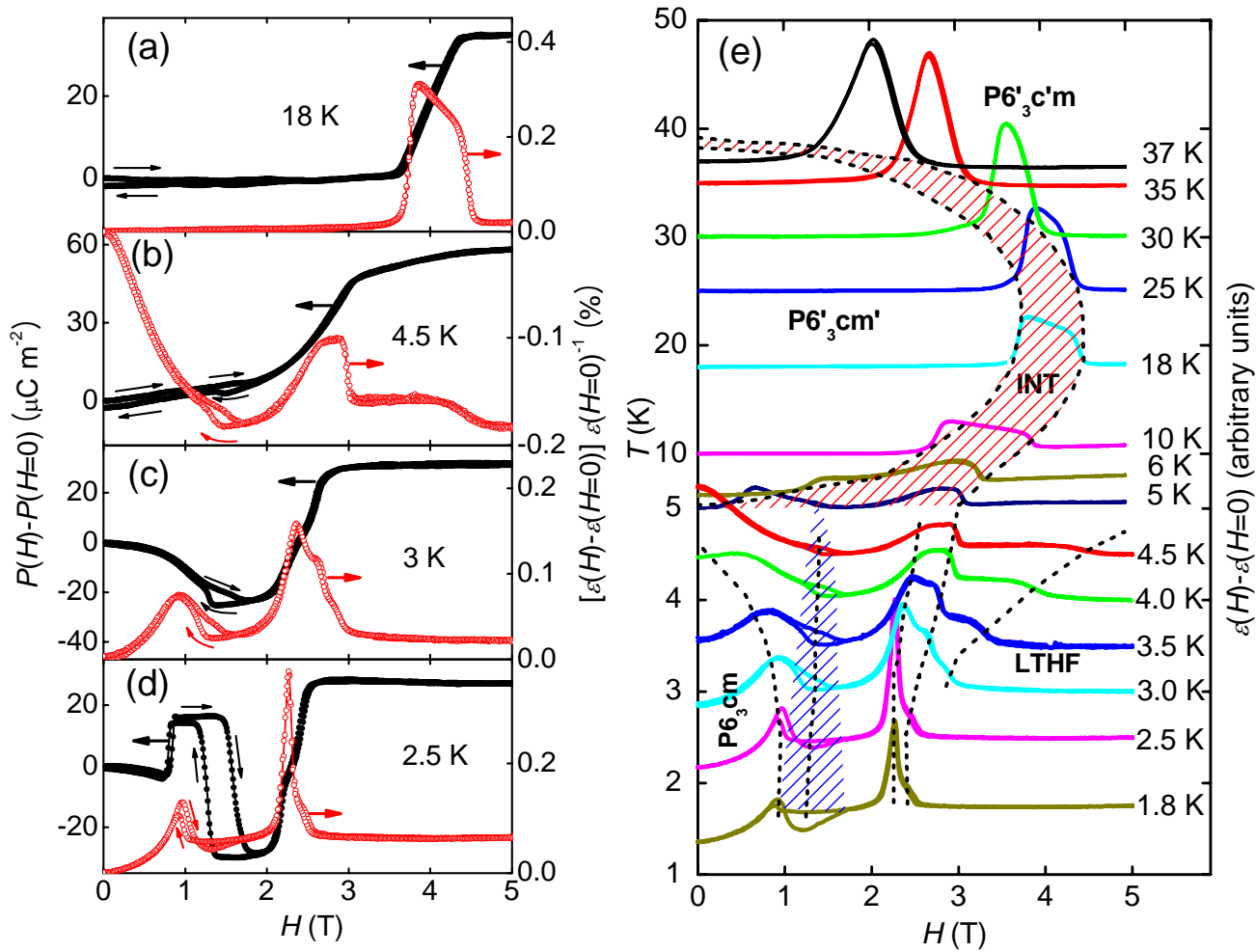

Fig. 2   N. Hur *et al.*

|  |  | x | y | z |
|---|---|---|---|---|
| Ho1 | 2a | 0 | 0 | 0.2747(5) |
| Ho2 | 4b | 1/3 | 2/3 | 0.2314(4) |
| Mn | 6c | 0.3359(9) | 0 | 0 |
| O1 | 6c | 0.3067(3) | 0 | 0.1631(3) |
| O2 | 6c | 0.6408(3) | 0 | 0.3366(4) |
| O3 | 2a | 0 | 0 | 0.4757(6) |
| O4 | 4b | 1/3 | 2/3 | 0.0175(4) |

TABLE 1    N. Hur *et al.*

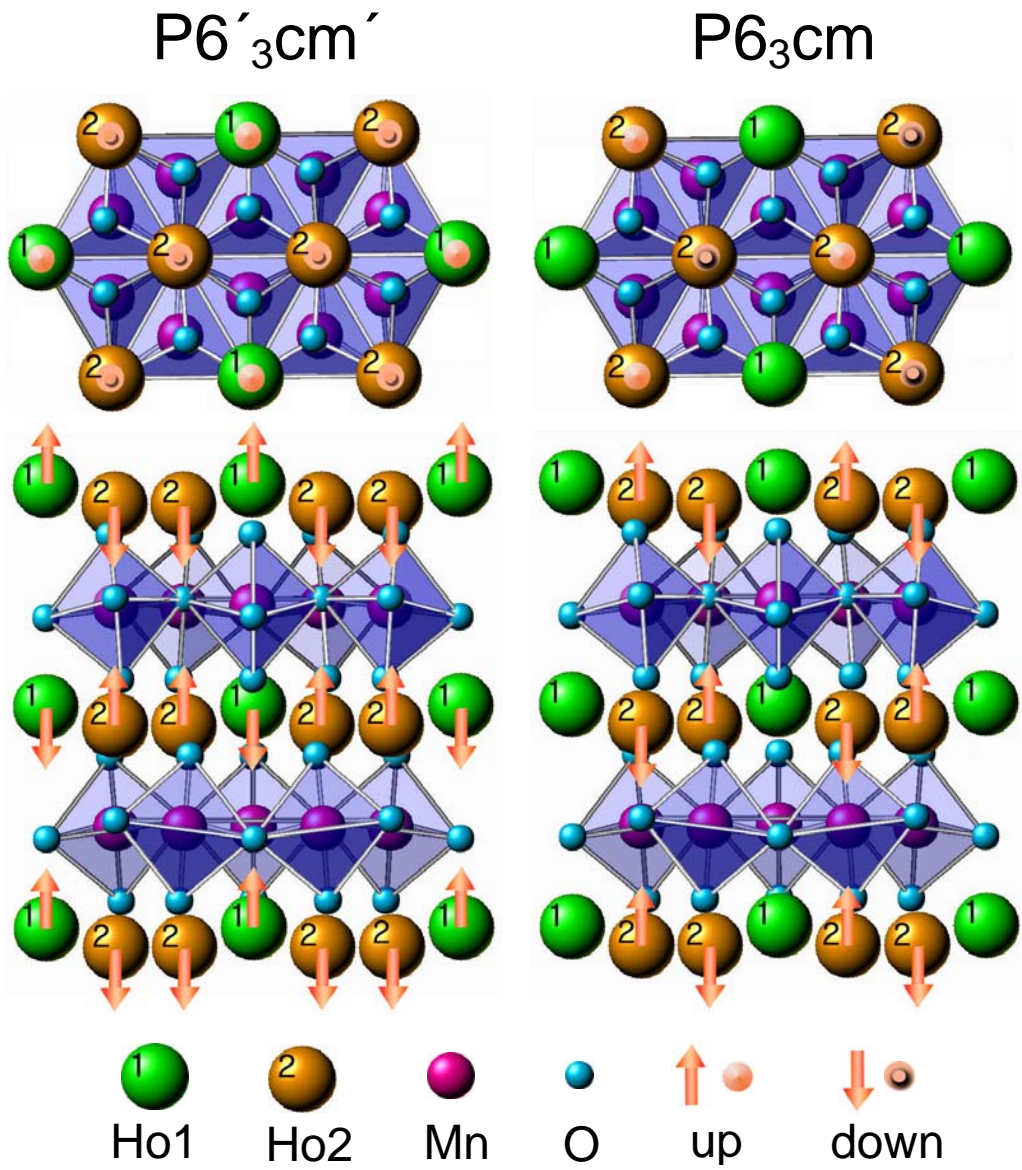

Fig. 3   N. Hur *et al.*